\documentclass[a4paper]{IEEEtran}
\usepackage[utf8]{inputenc}
\usepackage[]{algorithm2e}
\usepackage{color}
\usepackage{rotating}
\usepackage{amssymb}
\usepackage{amsmath}
\usepackage{graphicx}
\usepackage{float}
\usepackage{multirow}
\usepackage{longtable}
\usepackage{cite}
\usepackage{caption}
\usepackage{subfigure}

\makeatletter
\def\ps@IEEEtitlepagestyle{%
  \def\@oddfoot{\mycopyrightnotice}%
  \def\@evenfoot{}%
}
\def\mycopyrightnotice{%
  {\footnotesize 978-1-4799-6926-5/15/\$31.00 ©2015 IEEE.\hfill}%
}

\author{
	Huda Al-Nayyef$^{1, 2}$, Christophe Guyeux$^1$, and	Jacques M. Bahi$^1$\\
	$^1$~FEMTO-ST Institute, UMR 6174 CNRS, DISC Computer Science Department \\
	Universit\'e de Franche-Comt\'e,
	16, Rue de Gray, 25000 Besan\c{c}on, France\\
$^2$~Computer Science Department, University of Mustansiriyah, Iraq\\
\{huda.al-nayyef, christophe.guyeux, jacques.bahi\}@univ-fcomte.fr
}
\title{\emph{Taenia} Biomolecular Phylogeny and the Impact of Mitochondrial Genes on this Latter}

\begin{document}

\maketitle

\begin{abstract}
Variations in mitochondrial genes are usually considered to infer phylogenies. 
However some of these genes are lesser constraint than other ones, and thus
may blur the phylogenetic signals shared by the majority of the mitochondrial
DNA sequences. To investigate such effects,  
in this research work, the molecular phylogeny of the genus \emph{Taenia} is studied using
14 coding sequences extracted from mitochondrial genomes of 17 species. 
We constructed 16,384 trees, using a combination of 1 up to 14 genes. We obtained 131 topologies, and we showed that only four particular instances were relevant. Using further statistical investigations, we then extracted a particular topology, which displays more robustness properties. 

\end{abstract}

\begin{IEEEkeywords}
\emph{Taenia},
Phylogeny, 
Statistical tests
\end{IEEEkeywords}

\section{Introduction}

\emph{Taenia} (Cestoda: Taeniidae) is a genus of tapeworm (a type of helminth) members that have some important parasites of livestock. These parasitic organisms handle taeniasis and cysticercosis in humans, which are a type of helminthiasis that was belonging to the group of neglected tropical diseases \cite{roberts2005gerald}. Despite intensive research, the taxonomy of this genus remains unclear. Based on morphology and life cycle data. 
An essential key to solve the last issues raised by \emph{Taenia} is believed to be found in the study of the large amount of recently available DNA sequences, especially with complete mitochondrial (mt) genomes. Genes from mt genomes are classical markers for phylogeny. This DNA presents interesting features for such analysis: genes are shared by almost all eukaryotes and are present in a single copy, the molecule is maternally inherited and non-recombining in most cases, etc.~\cite{Ballard}.

Part of the problem resides in the fact that, even though the amount of information should be sufficient to infer a correct phylogeny of this genus. The presence of homoplasy in individually available genes clouds the general phylogenetic message, raising uncertainties in some locations of the tree. The question we discuss in the present work is thus to determine 
which genes are homoplastic, and which ones tell the story of the
species. Our goal is thus to exhibit a well-supported phylogenetic tree of the genus \emph{Taenia}. Our analysis relies on 
some recent statistical tools and intensive computations
on available bio-molecular data~\cite{aagcsap16:ij,acgmsb+14:ij}.

After a presentation of the major problems that remain need to be solved regarding the phylogeny of \emph{Taenia}, we will describe in details our investigation protocol. 
We will then present how each phylogenetic tree inference has been conducted. Our approach mainly based on annotating from scratch each genome, using an efficient alignment tool, and various mutation models for mitochondrial coding sequences and RNAs. We will
then explain how we have obtained the $16,384$ phylogenetic trees of that study, and how
we have used them to solve the phylogenetic reconstruction problem for this genus as a result of estimating the influence of each gene on that topology.
 
\begin{table}[]
\centering
\caption{\emph{Taenia} species analyzed in this paper and accession numbers of mitochondrial genomes. (\emph{E. vogeli}) is an outgroup.}
\begin{tabular}{ll}
\hline
Species  & Accession \\
\hline
\emph{Taenia asiatica} & NC\_004826\\
\emph{Taenia crassiceps} & NC\_002547\\
\emph{Taenia hydatigena} & NC\_012896\\
\emph{Taenia krepkogorski} & NC\_021142\\
\emph{Taenia laticollis} & NC\_021140\\
\emph{Taenia madoquae} & NC\_021139\\
\emph{Taenia martis} & NC\_020153\\
\emph{Taenia multiceps} & NC\_012894\\
\emph{Taenia mustelae} & NC\_021143\\
\emph{Taenia ovis} & NC\_021138\\
\emph{Taenia parva} & NC\_021141\\
\emph{Taenia pisiformis} & NC\_013844\\
\emph{Taenia saginata} & NC\_009938\\
\emph{Taenia serialis} &  NC\_021457\\
\emph{Taenia solium} & NC\_004022\\
\emph{Taenia taeniaeformis} & NC\_014768\\
\emph{Taenia twitchelli} & NC\_021093\\
\hline
\emph{Echinococcus vogeli} & NC\_009462\\
\hline
\end{tabular}
\label{listeGenomes}
\end{table}

\label{section:State}
To date, 17 complete mitochondrial genomes of \emph{Taenia} have been published, their list and accession number being provided in Table~\ref{listeGenomes}. These genomes have been used recently to update the phylogeny of this genus using molecular data. As presented in Table~\ref{listeGenomes2} many previous articles of phylogeny have worked with \emph{Taenia} species, but none of them provides a well-supported tree for this genus. For this reason, the authors of this paper proposed the new computational methods for constructing and finding a well-supported phylogenetic tree for \emph{Taenia}~\cite{aangc+15:ip}.

The remainder of this article is constituted as follows. Section~\ref{sec:methodo} is devoted to the proposed methodology intended to improve the estimation of the phylogenetic tree. Finer statistical investigations of the homoplastic character of certain genes are detailed in Section~\ref{sec:investigations}. This article ends with a discussion and a description of possible future work on this problem.
\begin{table}[]
\centering
\caption{\emph{Taenia} (\emph{Eucestoda}) in state of the art phylogenies (* when serving as outgroup; + when present in dataset but not used in phylogenetic analyses; $X^n$ when
$n$ representents of the species).}
\scriptsize
\begin{tabular}{lcccccccccc}
 & \begin{turn}{90}Hoberg \emph{et al.} $2000$~\cite{Hoberg00a}\end{turn} & \begin{turn}{90}Hoberg \emph{et al.} $2001$~\cite{Hoberg00} \end{turn}& \begin{turn}{90} Hoberg $2006$~\cite{Hoberg06} \end{turn} & \begin{turn}{90} Nakao \emph{et al.} $2010$~\cite{Nakao10} \end{turn} & \begin{turn}{90} Lavikainen \emph{et al.} $2010^\circ$~\cite{Lavikainen10} \end{turn} & \begin{turn}{90} Knapp \emph{et al.} $2011^\circ$~\cite{Knapp11} \end{turn} & \begin{turn}{90} Nakao \emph{et al.} $2013^\circ$~\cite{Nakao13} \end{turn} & \begin{turn}{90} $\textrm{This study}^\circ$  \end{turn}  & \begin{turn}{90}Total of studies  \end{turn} \\
\hline
\emph{Taenia acinonyxi}                  & X & X & X                    &                     &                &    &    &   & 3\\
\emph{Taenia asiatica}                   & X & X & X                    & X                   & X              & X  & X  & X & 8\\
\emph{Taenia brachyacantha}              & X & X &                      &                     &                &    &    &   & 2\\
\emph{Taenia crassiceps}                 & X & X & X                    & X                   & X              & X  & X  & X & 8\\
\emph{Taenia crocutae}                   & X & X & X                    & X                   &                &    &    &   & 4\\
\emph{Taenia dinniki}                    & X & X &                      &                     &                &    &    &   & 2\\
\emph{Taenia endothoracicus}             & X & X & X                    &                     &                &    &    &   & 3\\
\emph{Taenia gonyamai}                   & X & X & X                    &                     &                &    &    &   & 3\\
\emph{Taenia hyaenae}                    & X & X & X                    & X                   &                &    &    &   & 4\\
\emph{Taenia hydatigena}                 &   & X & X                    &                     & X              & X  & X  & X & 6\\
\emph{Taenia ingwei}                     & X & X &                      &                     &                &    &    &   & 2\\
\emph{Taenia intermedia}                 &   &   & X                    &                     &                &    &    &   & 1\\
\emph{Taenia krabbei}                    &   &   &                      &                     & X              &    &    &   & 1\\
\emph{Taenia krepkogorski}               &   &   &                      &                     &                &    & X  & X & 2\\
\emph{Taenia laticollis}                 & X & X &                      &                     &                & X  & X  & X & 5\\
\emph{Taenia macrocystis}                & X & X & X                    &                     &                &    &    &   & 3\\
\emph{Taenia madoquae}                   & X & X & X                    &                     & X              & X  & X  & X & 7\\
\emph{Taenia martis}                     & X & X & X                    &                     & X              & X  & X  & X & 7\\
\emph{Taenia multiceps}                  & X & X & X                    & X                   & X              & X  & X  & X & 8\\
\emph{Taenia mustelae}                   & X & X & X                    &                     & X              & X  & X  & X & 7\\
\emph{Taenia olngojinei}                 & X & X & X                    &                     &                &    &    &   & 3\\
\emph{Taenia omissa}                     & X & X & X                    &                     &                &    &    &   & 3\\
\emph{Taenia ovis}                       & X & X & $\textrm{X}^2$       &                     & X              & X  & X  & X & 7\\
\emph{Taenia parenchymatosa}             & X & X & $\textrm{X}^3$       &                     &                &    &    &   & 3\\
\emph{Taenia parva}                      & X & X & X                    &                     & X              & X  & X  & X & 7\\
\emph{Taenia pencei}                     &   &   & X                    &                     &                &    &    &   & 1\\
\emph{Taenia pisiformis}                 & X & X & X                    &                     & X              &    &    & X & 5\\
\emph{Taenia polyachantha}               & X & X &                      &                     & $\textrm{X}^2$ &    &    &   & 3\\
\emph{Taenia pseudolaticollis}           & X & X &                      &                     &                &    &    &   & 2\\
\emph{Taenia regis}                      & X & X & X                    &                     & X              &    &    &   & 4\\
\emph{Taenia rileyi}                     & X & X & X                    &                     &                &    &    &   & 3\\
\emph{Taenia saginata}                   & X & X & X                    & X                   & X              & X  & X  & X & 8\\
\emph{Taenia selousi}                    & X & X & X                    &                     &                &    &    &   & 3\\
\emph{Taenia serialis}                   & X & X & $\textrm{X}^2$       & $\textrm{X}^2$      & X              & X  & X  & X & 8\\
\emph{Taenia simbae}                     & X & X & X                    & X                   &                &    &    &   & 4\\
\emph{Taenia solium}                     & X & X & X                    & X                   & X              & X  & X  & X & 8\\
\emph{Taenia taeniaeformis}              & X & X & X                    & X                   & $\textrm{X}^2$ & X  & X  & X & 8\\
\emph{Taenia taxidiensis}                & X & X & X                    &                     &                &    &    &   & 3\\
\emph{Taenia twitchelli}                 & X & X & X                    &                     & X              & X  & X  & X & 7\\
\emph{Echinococcus vogeli}               &   &   &                      & X                   &                & X  & +  &  * & 3\\
\hline
Total 40                                 &34 &35 & 31                   & 11                  & 18             & 16 & 16 & 17 &\\
\hline
\end{tabular}
\label{listeGenomes2}
\end{table}

\section{Materials and methods}
\label{sec:methodo}

\subsection{Alignment and annotations of coding sequences}

\begin{table}[]
\centering
\caption{Details of obtained topologies. The lowest bootstrap and the number of occurrence of each calculated topology is indicated.}
\scalebox{0.7}{
\begin{tabular}{|c|c|c|c|l|}
\hline
Topology & Lowest    & Number of   & Average   & Discarded \\
         & bootstrap & occurrences & bootstrap & genes\\
\hline
0 & 82 & 2049 & 44 & \emph{Atp6, Cob, Cox2, Nad1, Nad2, Nad3, Nad5}  \\
1 & 84 & 6442 & 51 & \emph{Nad1, Nad3, Nad5, Nad6, Rrns} \\
2 & 92 & 3276 & 52 & \emph{Cox2, Cox3, Nad4, Nad4l, Nad5, Rrnl, Rrns}  \\
3 & 76 & 931 & 48 & \emph{Atp6, Cox1, Nad1, Nad3, Nad4, Rrnl}  \\
4 & 74 & 452 & 52 & \emph{Atp6, Cob, Cox1, Cox3, Nad4, Nad5, Rrnl}  \\
5 & 56 & 317 & 28 & \emph{Cob, Cox1, Cox2, Cox3, Nad1, Nad2, Nad3, Nad4l, Rrnl, Rrns}  \\
6 & 68 & 614 & 39 & \emph{Atp6, Cox1, Cox2, Cox3, Nad2, Nad3, Nad5}  \\
7 & 68 & 321 & 43 & \emph{Atp6, Cox2, Cox3, Nad1, Nad2, Nad3, Nad4, Nad4l, Nad6, Rrns}  \\
8 & 70 & 226 & 46 & \emph{Cob, Cox1, Cox2, Cox3, Nad4, Nad4l}  \\
9 & 58 & 69 & 39 & \emph{Cox1, Cox2, Cox3, Nad1, Nad3, Nad4, Rrns}  \\
10 & 74 & 230 & 45 & \emph{Atp6, Cob, Cox1, Nad1, Nad2, Nad4, Nad4l, Nad6, Rrnl}  \\
11 & 76 & 172 & 53 & \emph{Cob, Cox1, Cox2, Cox3, Nad1, Nad3, Nad4, Nad5, Rrnl}  \\
12 & 60 & 212 & 30 & \emph{Atp6, Cox2, Cox3, Nad1, Nad2, Nad4l, Nad6, Rrns}  \\
13 & 56 & 92 & 42 & \emph{Atp6, Cob, Cox1, Cox2, Cox3, Nad1, Nad3, Nad4}  \\
14 & 64 & 39 & 44 & \emph{Atp6, Cob, Cox1, Cox2, Nad3, Nad4, Nad5, Nad6, Rrns}  \\ \hline
\end{tabular}}
\label{tab:infoTree}
\end{table}

To answer the aforementioned questions, first Bayesian and maximum likelihood analyses have been realized on either the whole mitogenomes or its twelve protein coding genes. Theses analyses were realized using nucleotides and translated amino acids sequences. Tools used during these first runs of analyses were:
\begin{itemize}
\item Muscle~\cite{Muscle} for aligning complete mitogenomes and T-Coffee~\cite{tcoffee} for genes alignments;
\item NCBI annotations for coding sequences in a first analysis, and then DOGMA~\cite{DOGMA} in a deeper stage;
\item PhyloBayes~\cite{phylobayes} for Bayesian inference, while PhyML~\cite{phyml} and RAxML~\cite{RAxML} have been used for maximum likelihood.
\end{itemize}

At each time, a problem of support (at least one bootstrap lower than 95, 
while a commonly accepted rule claims that all supports must be larger than this threshold~\cite{felsenstein1985confidence}) was found at least at one location of the obtained tree. Partial 
conclusions of these preliminary studies were that: (1) to use coding sequences is better than to 
consider the whole mitogenome, (2) there are inconsistencies in NCBI annotations, (3) T-Coffee alignments 
seem better than muscle ones, (4) many coding sequences narrate the story of the genus while others  tell their own history, and (5) to enlarge the amount of data leads to 
more supported trees.

\subsection{Methodological approach}

\begin{figure*}
\centering
\subfigure[Topology 0]{\includegraphics[scale=0.45]{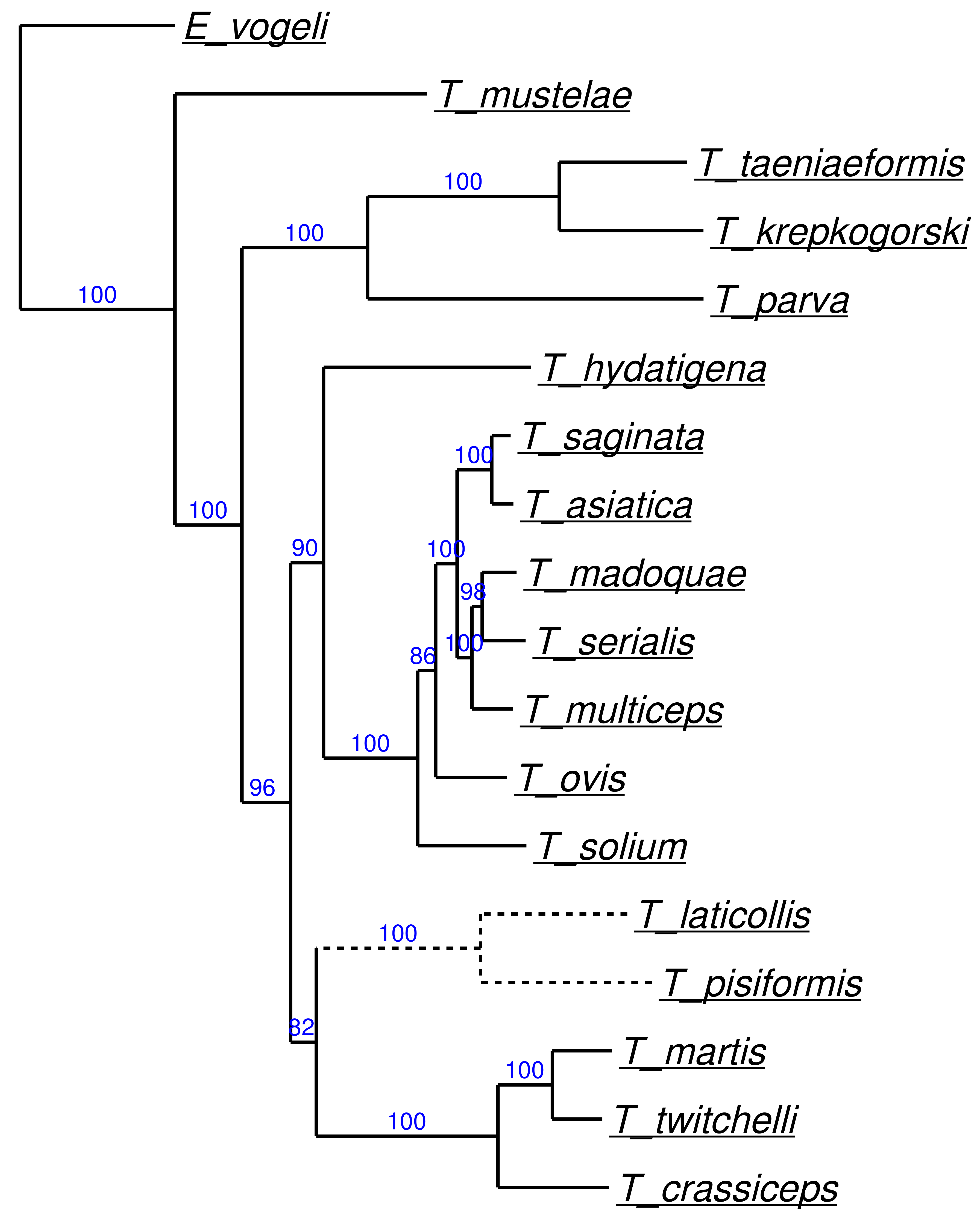}\label{bestTopoEchino}}
\subfigure[Topology 1]{\includegraphics[scale=0.45]{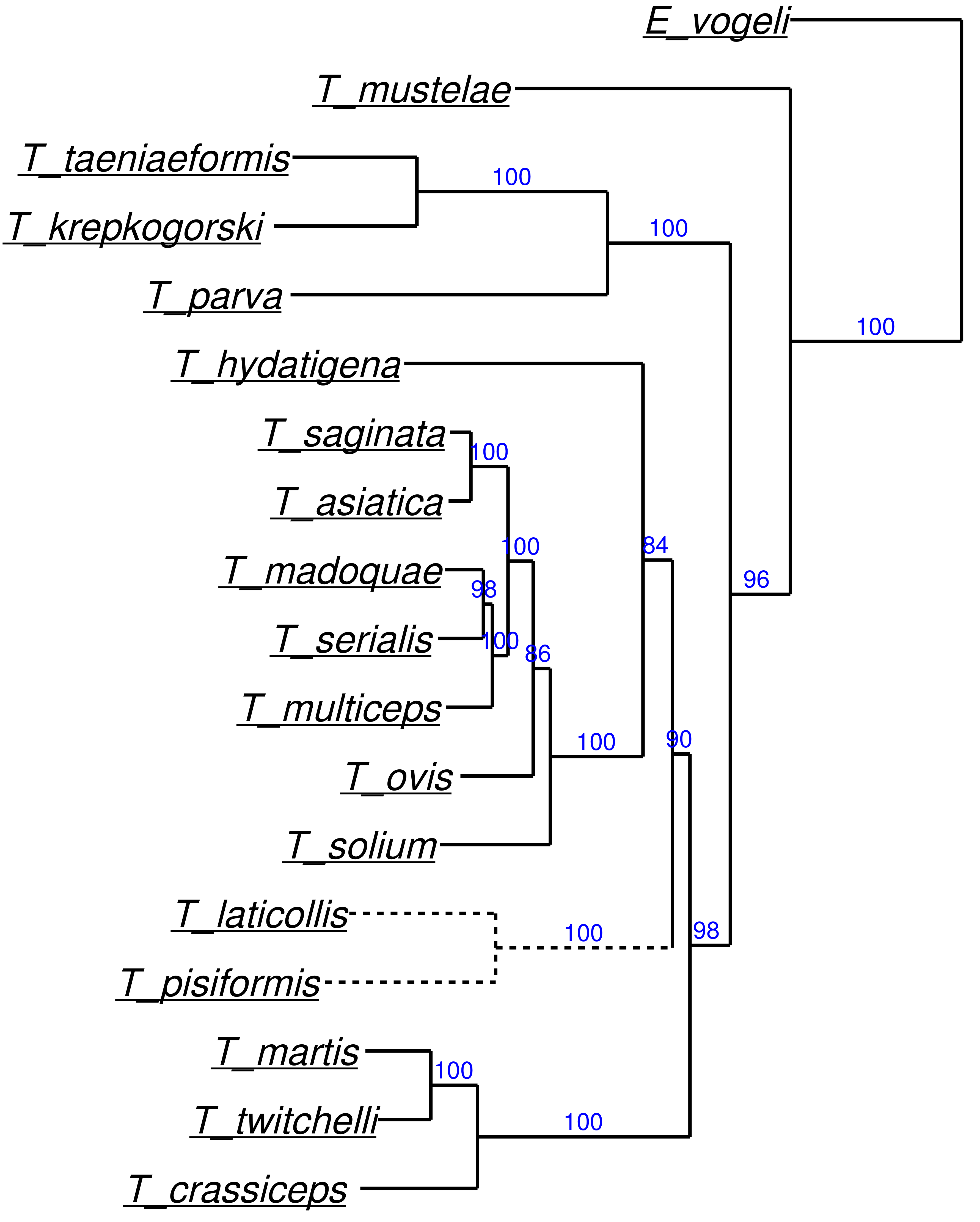}}
\subfigure[Topology 2]{\includegraphics[scale=0.45]{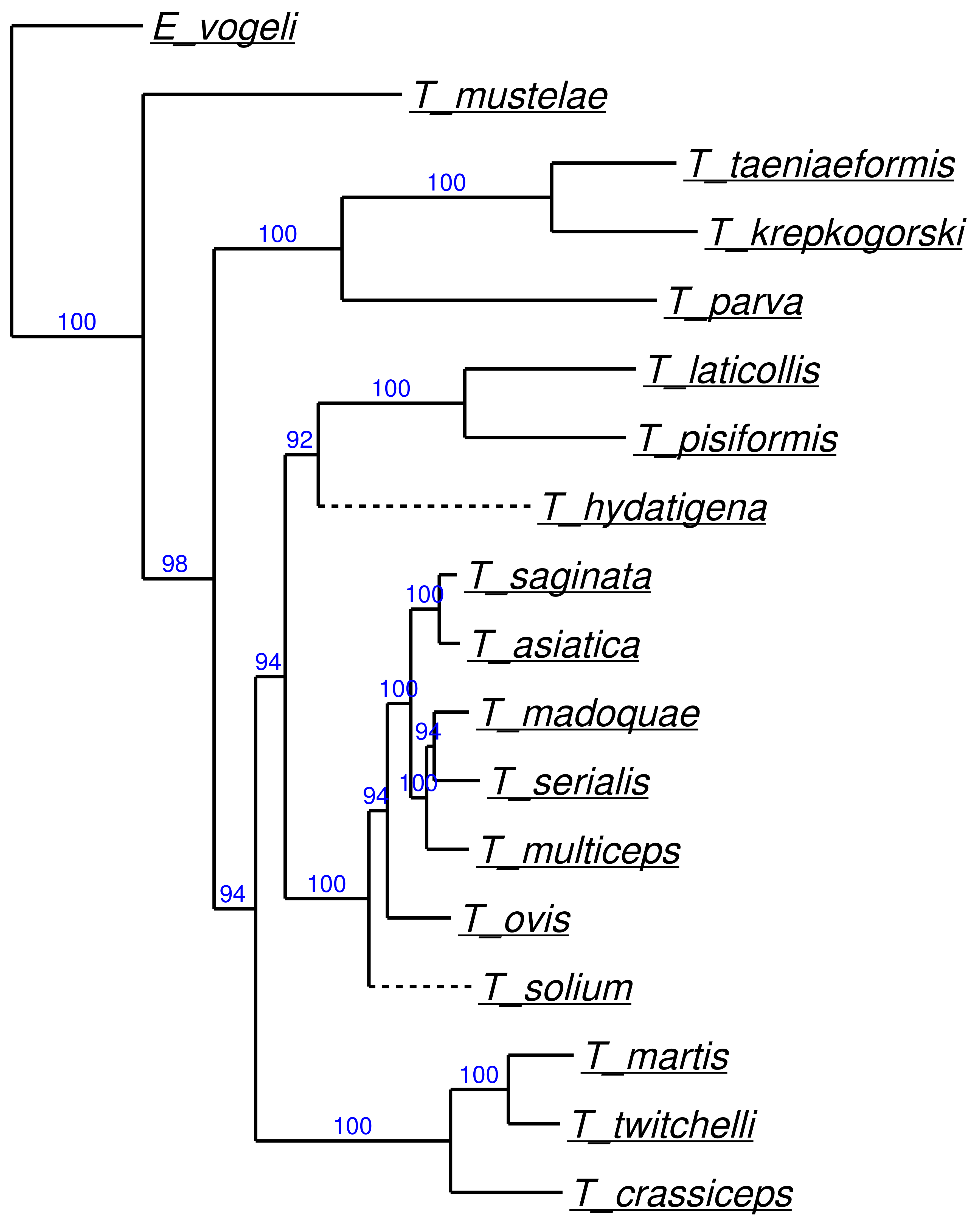}}
\subfigure[Topology 3]{\includegraphics[scale=0.45]{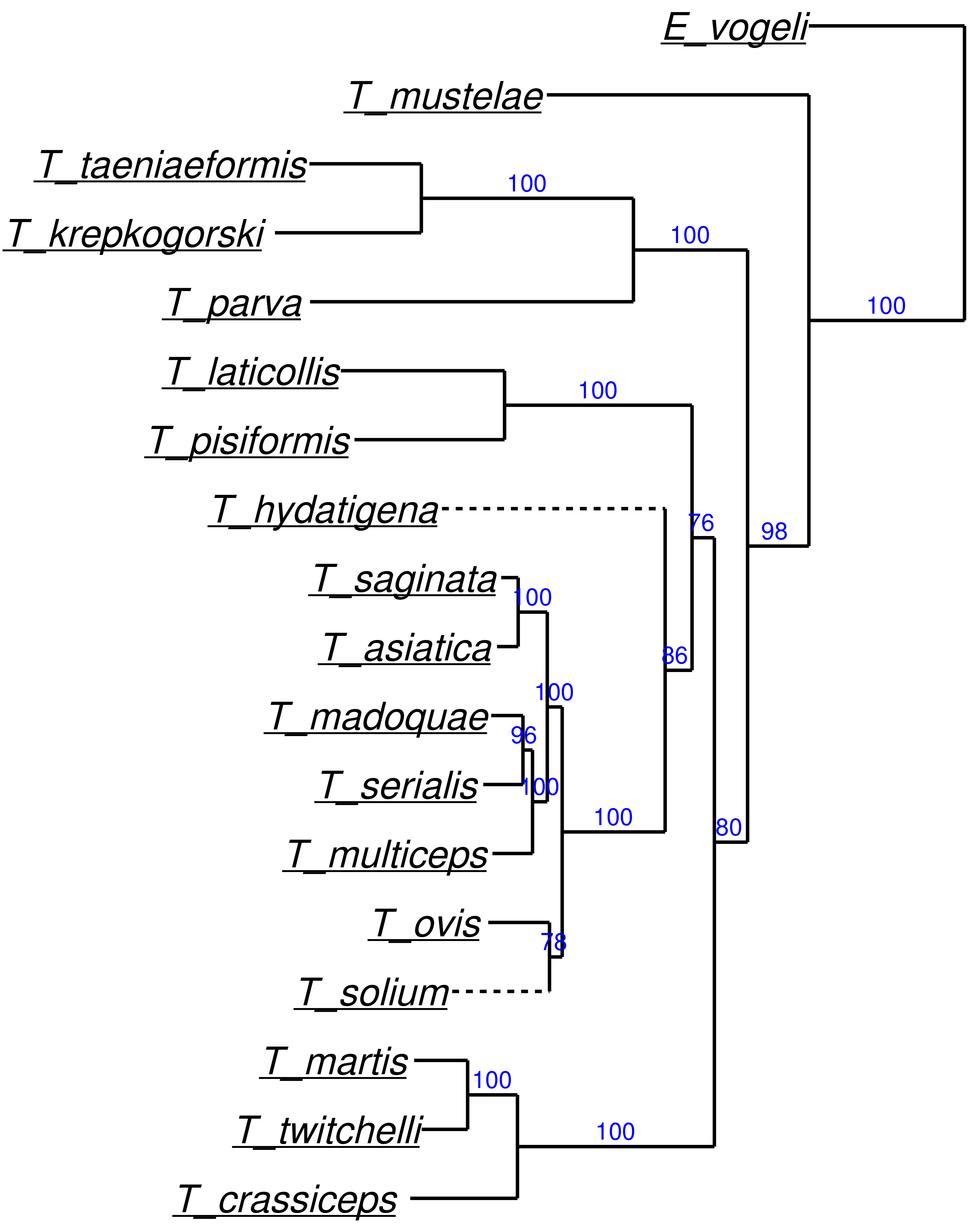}}
\caption{4  trees with more than 700 occurrences, when considering 16,384 trees obtained with Algorithm~\ref{algo1} (\emph{E. vogeli} is an outgroup)}
\label{fig:lesBestTopo}
\end{figure*}

To solve both the phylogeny of \emph{Taenia} and the determination of genes that break it, 
a solution has been to consider all available or obtainable coding sequences shared by these 18 species, and to 
investigate how the inferred phylogenies evolve when using a various subset of these sequences. Doing so enlarge the first investigations of Hardman \emph{et al.}~\cite{ZSC:ZSC248}, who have studied the phylogeny of 5 \emph{Taeniidae} according to each of the 12 mitochondrial genes taken alone, 14 sequences have been extracted from each of the considered species: 12 protein coding sequences and 2 
rRNAs from the mitochondrial genomes. They are listed below.

\begin{itemize}

\item Mitochondrial protein coding sequences:

\emph{atp6} (ATP synthase 6), \emph{cob} (cytochrome b), \emph{cox1} (cytochrome c oxydase 1), \emph{cox2} (cytochrome c oxydase 2), \emph{cox3} (cytochrome c oxydase 3), \emph{nad1} (NADH dehydrogenase subunit 1), \emph{nad2} (NADH dehydrogenase subunit 2), \emph{nad3} (NADH dehydrogenase  subunit 3), \emph{nad4} (NADH dehydrogenase 4), \emph{nad4l} (NADH dehydrogenase subunit 4L), \emph{nad5} ((NADH dehydrogenase subunit 5), \emph{nad6} (NADH dehydrogenase subunit 6).
\item Mitochondrial rRNAs:

\emph{rrnL} (large subunit rRNA), \emph{rrnS} (small subunit rRNA).
\end{itemize}
DOGMA, for its part, has been used to annotate from scratch each up-to-date complete mitochondrial genome downloaded from NCBI \cite{Genbank2008} 
Default parameters of DOGMA have been selected, namely an identity cutoff for protein equal to $60\%$ and $80\%$ for coding genes and rRNAs respectively for \emph{Taenia} species, while these thresholds have been reduced to $55\%$ and $75\%$ for \emph{T. mustelae}, due to a problem of detection of \emph{nad6} and \emph{rrnL} respectively.
The e-value was equal to $1e-5$, and the number of blast hits to return has been set to 5. 

Each of these 14 coding sequences has been aligned separately by using T-Coffee (M-Coffee mode, using 6 cores for multiprocessing). Then 
16,384 trees were constructed, corresponding to all the possible combinations of 1, 2, 3, ..., and 14 coding
sequences among the 14 available ones ($\sum_{k=1}^{14} \binom{k}{14} = 16,384$), as described in Algorithm~\ref{algo1}. 
This computation has taken 3 months on the ``Mésocentre de Calcul de Franche-Comté'' supercomputer facilities.
The idea behind was to determine 
both the most supported phylogenetic trees and the effects of each gene on topologies and supports.
RAxML version 8.0.20 were used for maximum likelihood inference, with 3 distinct models/data partitions with joint branch length optimization at each computation, corresponding to the mitochondrial rRNAs, and the mitochondrial protein coding sequences. All free model parameters have been estimated by RAxML for both GAMMA model of rate heterogeneity and ML estimate of alpha-parameter. At each time, a maximum of 1000 non-parametric bootstrap inferences was executed, with MRE-based bootstopping criterion, and \emph{E. vogeli}
has been used as outgroup.

\begin{algorithm}[h]
 \For{k=1,...,14}{
 \For{each combination c of k genes}{
  build a phylogenetic tree T based on these k genes\;
  extract the list of bootstraps L(c) and the topology T(c)\;
  store (c, L(c), T(c))\;
 }
}
 \caption{Pseudocode producing $16,384$ phylogenetic trees}
 \label{algo1}
\end{algorithm}

\section{Discussion and results}
\label{sec:investigations}
\subsection{Results}
\begin{figure}
\centering
\subfigure[Number of trees per topology.]{\includegraphics[scale=0.32]{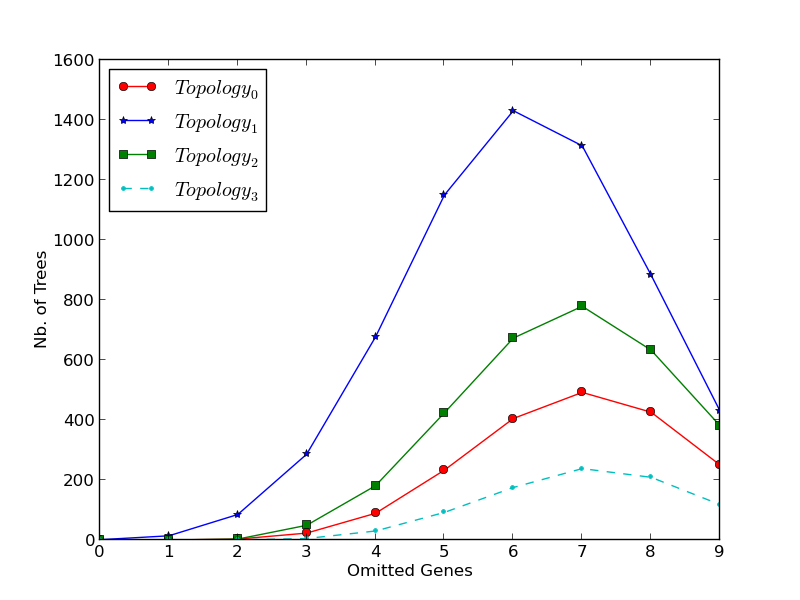}\label{fig:treesPerTopology}}
\subfigure[Number of trees whose lowest bootstrap is larger than 70.]{\includegraphics[scale=0.32]{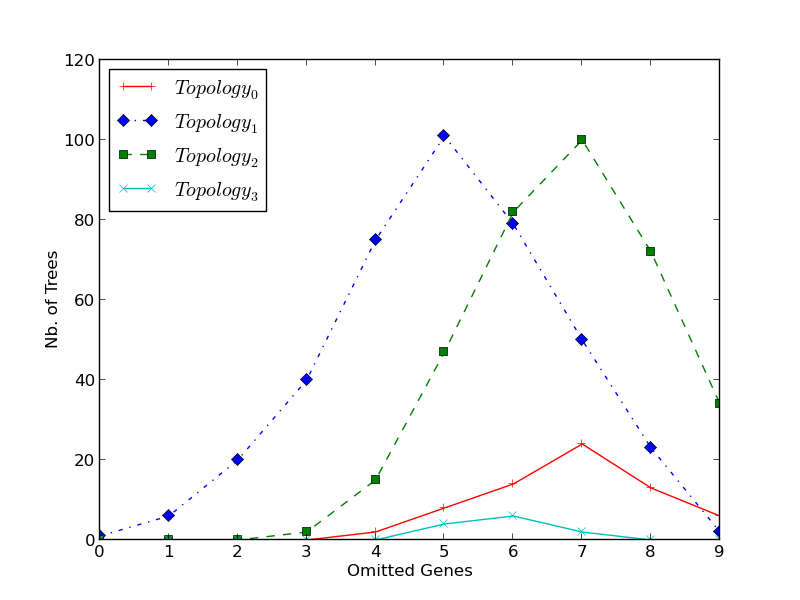}\label{fig:numberOfGood}}\\
\subfigure[Lowest bootstrap in best trees.]{\includegraphics[scale=0.32]{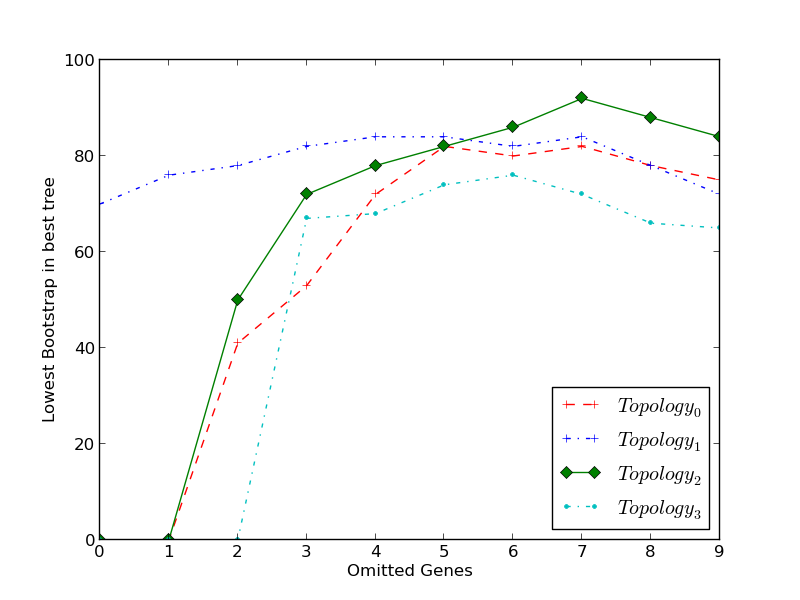}\label{fig:treesPerLowestBootstrap}}
\subfigure[Average value of lowest bootstraps]{\includegraphics[scale=0.32]{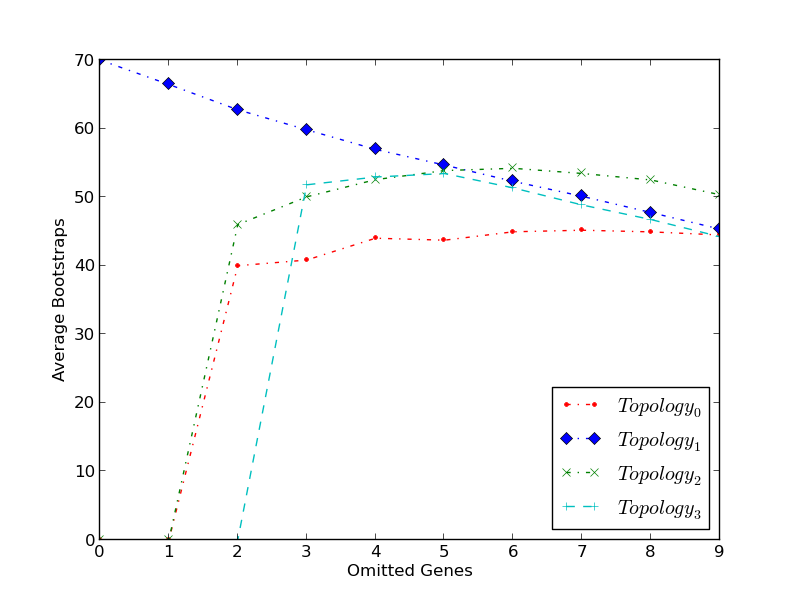}}
\caption{Comparison of the 4 best topologies, according to the number of discarded genes. A. The number of trees in each topology, according to the number of discarded genes. B. As in (a), but by considering only trees whose supports are larger than 70. C. Minimal support in the best tree of each topology, when regarding the number of discarded genes. D. Average of all minimum bootstrap in each tree of each topology.}
\label{fig:my_label}
\end{figure}

131 topologies were obtained during our computations 
with 17 species and 1 outgroup. 
Further information regarding these trees are provided in Table~\ref{tab:infoTree}: in this latter, we investigated the 15 most frequent topologies that contained 15,442 of the 16,384 trees (94.16\%). 
For each topology, the lowest bootstrap of the best tree (that is, the lowest bootstrap of the tree that maximizes the minimum taken over all its bootstraps), the number of trees having this topology, the average minimal bootstrap value, and the list of genes that have been removed to obtain the best tree having this topology, are provided. Only 4 of these 131 topologies 
have a number of occurrences larger than 700, when considering the 16,384 obtained trees.
They are depicted in Figure~\ref{fig:lesBestTopo}. 

These 4 best topologies representing 77.07\% of the obtained trees share most of their structure. For instance, \emph{T.~madoquae}, \emph{T.~serialis},  \emph{T.~multiceps},
are within a same clade, which is sister to the clade consisting of \emph{T.~asiatica} and \emph{T.~saginata}.
The differences between these most frequent topologies are depicted with dotted lines in Figure~\ref{fig:lesBestTopo}
while Table~\ref{difference} summarizes them using CompPhy tool~\cite{fiorini2014compphy}.

\begin{table}[]
\centering
\caption{Distance between trees (\emph{i.e.}, symmetrical Robinson–Foulds distance) and the different species between topologies after removing the maximum agreement subtree consensus.} 

\begin{tabular}{|c|c|c|c|c|}
\hline
Differences & \textbf{Top. 0} & \textbf{Top. 1} & \textbf{Top. 2} & \textbf{Top. 3} \\ \hline
\textbf{Top. 0} & \_ & \shortstack{T\_laticollis\\T\_pisiformis\\(RF=2)} & \shortstack{T\_laticollis\\ T\_pisiformis\\(RF= 4)} & \shortstack{T\_laticollis\\T\_pisiformis\\T\_solium\\(RF= 4)}\\ \hline
\textbf{Top. 1} & \textbf{\_} & \textbf{\_} & \shortstack{T\_hydatigena\\(RF= 2)} & \shortstack{T\_solium\\(RF= 2)}\\ \hline
\textbf{Top. 2} & \_ & \_ & \_ & \shortstack{T\_hydatigena\\T\_solium\\(RF= 4)}\\ \hline
\textbf{Top. 3} & \_ & \_ & \_ & \_ \\ \hline
\end{tabular}
\label{difference}
\end{table}

Various reasons have led us to consider the Topology 1 depicted in Figure~\ref{bestTopoEchino} as the most probable one.
Firstly, this is the most frequent topology, representing 39.31\% of the produced trees while the second one (Topology 2) 
represents only 19.99\% of the trees. 
Furthermore, this topology remains the most frequent one when we focus on trees generated 
by removing 0, 1, 2, 4, 5, 6, and 7
genes (notice that the largest number of trees are obtained when removing 6 genes for Topology 1, while we need to discard 7 genes to reach the largest populations of trees for Topologies 0, 2, and 3, see Figure~\ref{fig:treesPerTopology}). 

The number of trees whose lowest bootstraps is greater than 70 is nearly the same for Topologies~1 and 2 but, in Topo.~1, the largest number of trees is obtained when 5 genes are discarded, while we need to remove 7 genes to reach this maximum with topology number 2, as depicted in Figure~\ref{fig:numberOfGood}. 
Additionally, we can remark that in the topology number 1, the lowest bootstrap
in the best tree does not evolve so much when removing between 0 and 7 genes (it ranges between 70 and 84) while in Topology~2, the best lowest value (92) in the best trees is obtained  with 7 gene loss (see Figure \ref{fig:treesPerLowestBootstrap}). 

\begin{figure}
\centering
\subfigure[Supertree for the whole 16,384 trees]{\includegraphics[scale=0.4]{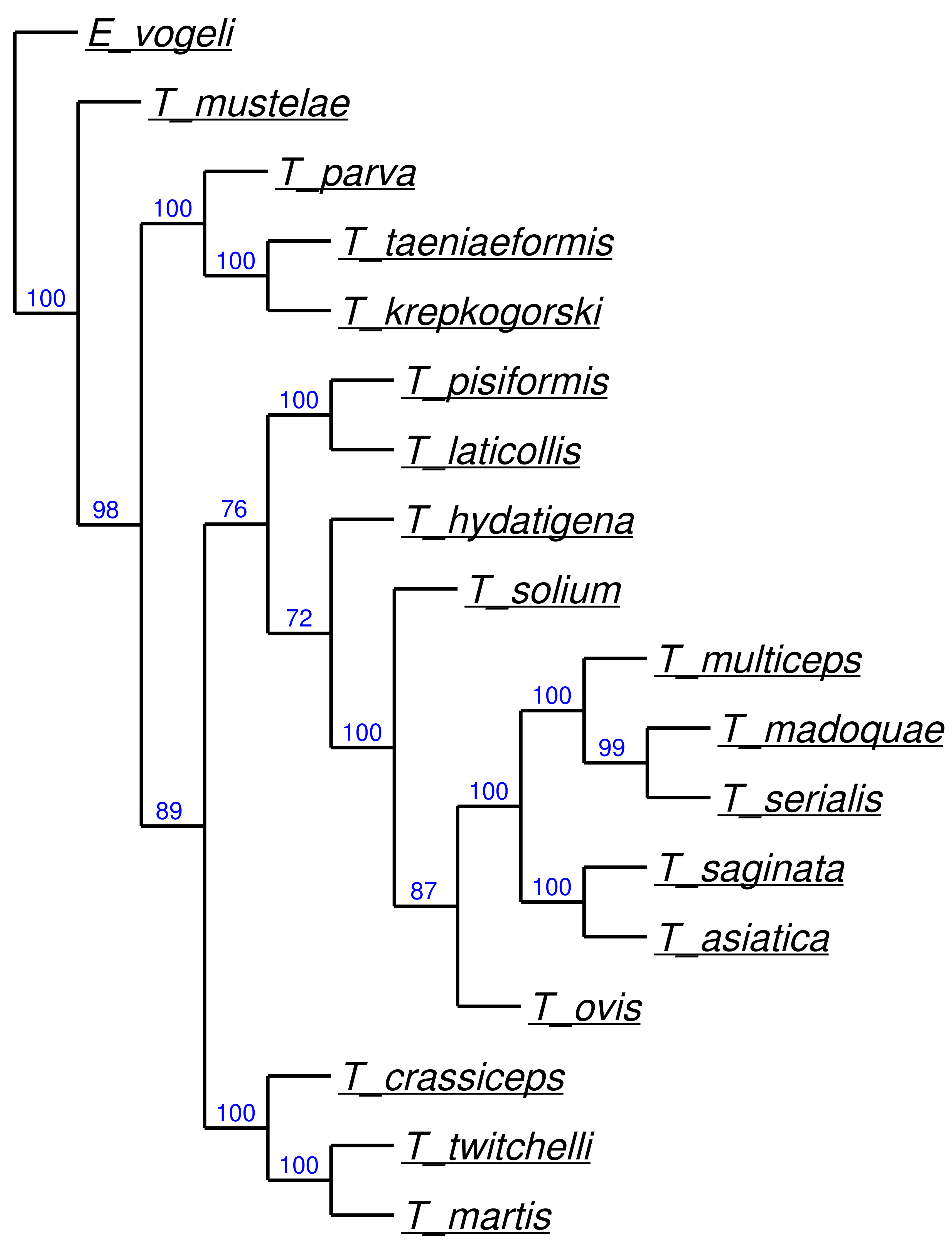}\label{fig:super4}}
\subfigure[Supertree for the trees of the 4 most frequent topologies]{\includegraphics[scale=0.4]{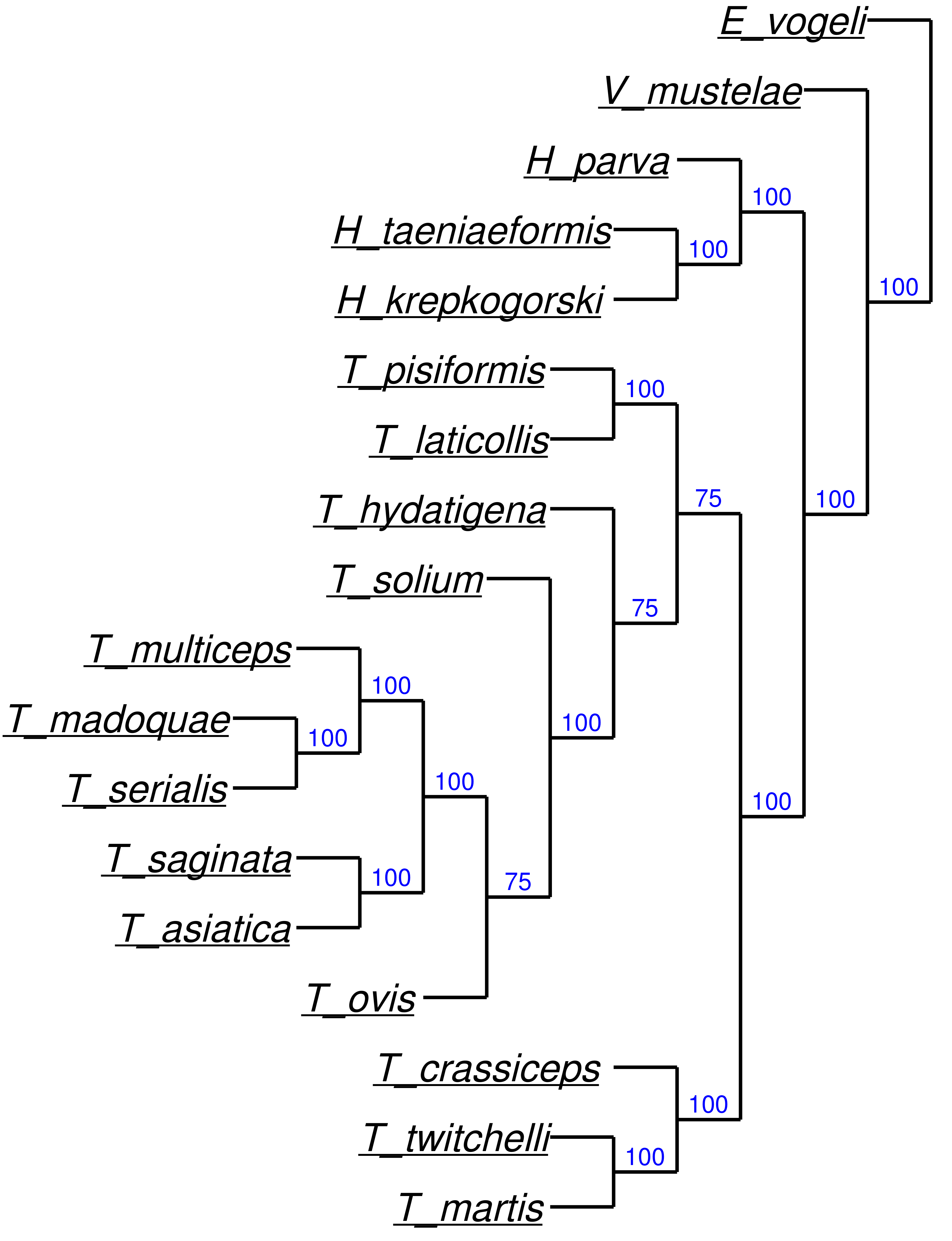}\label{fig:superall}}
\caption{Comparison of the supertrees obtained by SuperTriplets}
\label{fig:compsuper}
\end{figure}

Almost all results listed above tend to prove that Topology~1 is the best 
candidate for reflecting the \emph{Taenia} phylogeny. However
the fact that the tree having the best lowest bootstrap (82) belongs to Topology
number 2 raises certain questions. It is true that this latter has been obtained
by removing half of the genes, but there is no denial in the fact that topology
the most frequent and topology having the most supported tree are not the same.
To go deeper in the analysis of these topologies, we began to use SuperTriplets 
tool~\cite{Ranwez15062010} on the following two experiments. The 
supertree of all trees belonging to the four topologies presented before 
has been firstly computed, while in a second run
of experiments, the supertree for all 16,384 
phylogeny trees have been determined. Obtained results are reproduced in 
Figures~\ref{fig:super4} and~\ref{fig:superall}: at each time, Topology~1 has been obtained, thus reinforcing the view that this topology should reflect well the
Evolution history of \emph{Taenia}.

To validate this choice, next subsections will now investigate more deeply the 
relation between genes on the one hand, and both tree topologies and supports on 
the other hand.

\subsection{Gene occurrences}
\begin{algorithm}[h]
 \For{each gene g}{
 \For{each topology t}{
   count[g][t] = 0\;
 }
}
 \For{each (c,L,T) in the list stored by Algorithm~\ref{algo1}}{
 \For{each g in c}{
   count[g][t] = count[g][t]+1\;
 }
}
\caption{Pseudocode producing Table~\ref{tab:frequences}}
 \label{algo2}
\end{algorithm}

A first investigation consists of 
regarding if the presence of each gene is uniformly distributed in each of the 4 most frequent topologies, using Algorithm~\ref{algo2}.
Since each of the 16,384 produced trees is constructed using a subset of the 14 available sequences, it
is relevant to count the number of 
occurrences for each of these sequences. Table~\ref{tab:frequences} summarizes these results.

\begin{table}[h]
\centering
\caption{Number of times each gene were present to produce a tree having one of the 4 most relevant topologies.}
\scalebox{0.8}{
\begin{tabular}{|c||c|c||c|c||c|c||c|c|}
\hline
Topologies &  \multicolumn{2}{c||}{0} &  \multicolumn{2}{c||}{1} &  \multicolumn{2}{c||}{2} &  \multicolumn{2}{c|}{3} \\
            \hline
& number & rank & number & rank & number & rank & number & rank  \\
\hline
\hline
\emph{atp6}  & 924 & 9 & 3431 & 8 & 1924 & 4 & 423 & 9 \\ \hline
\emph{cob}  & 787 & 11 & \textit{\textbf{4179}} & \textit{\textbf{2}} & 1691 & 6 & 350 & 11 \\ \hline
\emph{cox1}  & 1209 & 4 & \textit{\textbf{4324}} & \textit{\textbf{1}} & 1326 & 11 & 542 & 5 \\ \hline
\emph{cox2}  & 1152 & 6 & 3740 & 6 & 1472 & 8 & \textit{\textbf{686}} & \textit{\textbf{2}} \\ \hline
\emph{cox3}  & 1469 & 3 & 3966 & 4 & \textit{\textbf{1260}} & \textit{\textbf{13}} & 449 & 7 \\ \hline
\emph{nad1}  & \textit{\textbf{584}} & \textit{\textbf{13}} & 3379 & 12 & \textit{\textbf{2549}} & \textit{\textbf{2}} & 257 & 12 \\ \hline
\emph{nad2}  & \textit{\textbf{84}} & \textit{\textbf{14}} & 3391 & 11 & \textit{\textbf{3004}} & \textit{\textbf{1}} & 527 & 6 \\ \hline
\emph{nad3}  & 1142 & 7 & \textit{\textbf{2708}} & \textit{\textbf{13}} & 2069 & 3 & 448 & 8 \\ \hline
\emph{nad4}  & \textit{\textbf{1699}} & \textit{\textbf{1}} & 3677 & 7 & 1339 & 10 & \textit{\textbf{84}} & \textit{\textbf{13}} \\ \hline
\emph{nad4l}  & 1153 & 5 & 3421 & 10 & 1291 & 12 & 624 & 3 \\ \hline
\emph{nad5}  & 613 & 12 & 4139 & 3 & \textit{\textbf{694}} & \textit{\textbf{14}} & \textit{\textbf{883}} & \textit{\textbf{1}} \\ \hline
\emph{nad6}  & 937 & 8 & 3421 & 9 & 1887 & 5 & 390 & 10 \\ \hline
\emph{rrnL}  & 858 & 10 & 3855 & 5 & 1583 & 7 & \textit{\textbf{63}} & \textit{\textbf{14}} \\ \hline
\emph{rrnS}  & \textit{\textbf{1638}} & \textit{\textbf{2}} & \textit{\textbf{2598}} & \textit{\textbf{14}} & 1392 & 9 & 603 & 4 \\ \hline
\end{tabular}}
\label{tab:frequences}
\end{table}

A correlation seems to appear between some genes, either over or under-represented, and some topologies. More precisely, the following information
can be deduced by checking the effects of the three least frequent genes:
\begin{enumerate}
\item \emph{nad1} is ranked as the least frequent gene in Topologies 0, 1, and 3, while this gene is the second most frequent one for Topology~2 (\emph{i.e.}, this mitochondrial coding sequence gene plays an essential role in Topology~2).
\item It seems that taking \emph{nad5} into consideration leads to a move of \emph{T.~hydatigena} 
in the tree, as it is ranked 12 and 14 for Topologies 0 and 2 respectively, and 3 and 1 for Topologies 1 and 3 respectively.
\item Similarly, \emph{rrnS}
seems responsible for the position evolution of \emph{T.~laticollis} and \emph{T.~pisiformis}: this gene is ranked in 2nd position for Topology~0 while this is the least frequent gene (last position) for Topology~1.
\item Gene \emph{nad5} is ranked first for Topology~3, so it may impact the sister relationship between \emph{T.~solium} and \emph{T.~ovis}. 
\end{enumerate}
However, these claims need to be further investigated by a more rigorous statistical approach, which is the aim of the following
sections.

\subsection{Genes influence on topology using Dummy logit model}

\begin{table}[]
\centering
\caption{Dummy logit regression results for Topology 1}
\begin{tabular}{c|c|c|c|c|c}
& coef &   std err      &    z      & $P>|z|$  &  [95.0\% Conf. Int.]\\
\hline
\textit{atp6 } & -0.2412 & 0.034 & -7.06 & 0.000 & [-0.308, -0.174] \\ 
\textit{cob } & 0.6861 & 0.035 & 19.871 & 0.000 & [0.618, 0.754] \\ 
\textit{cox1 } & 0.8592 & 0.035 & 24.733 & 0.000 & [0.791, 0.927] \\ 
\textit{cox2 } & 0.1444 & 0.034 & 4.231 & 0.000 & [0.078, 0.211] \\ 
\textit{cox3 } & 0.4261 & 0.034 & 12.431 & 0.000 & [0.359, 0.493] \\ 
\textit{nad1 } & -0.3059 & 0.034 & -8.944 & 0.000 & [-0.373, -0.239] \\ 
\textit{nad2 } & -0.2915 & 0.034 & -8.526 & 0.000 & [-0.359, -0.224] \\ 
\textit{nad3 } & \textbf{-1.1113} & 0.035 & -31.673 & 0.000 & [-1.18, -1.042] \\ 
\textit{nad4 } & 0.0658 & 0.034 & 1.928 & 0.054 & [-0.001, 0.133] \\ 
\textit{nad4l } & -0.2532 & 0.034 & -7.409 & 0.000 & [-0.32, -0.186] \\ 
\textit{nad5 } & 0.6381 & 0.034 & 18.512 & 0.000 & [0.571, 0.706] \\ 
\textit{nad6 } & -0.2537 & 0.034 & -7.423 & 0.000 & [-0.321, -0.187] \\ 
\textit{rrnL } & 0.2873 & 0.034 & 8.403 & 0.000 & [0.22, 0.354] \\ 
\textit{rrnS } & \textbf{-1.2345} & 0.035 & -35.003 & 0.000 & [-1.304, -1.165] \\ 
\end{tabular}
\label{tab:logit}
\end{table}

To investigate more deeply the effects of each coding sequence on the species topology, 4
dummy binary choice logit models have been 
realized (one per each best topology) using \texttt{scikit-learn}~\cite{scikit-learn} module of Python language. The reference to the exogenous 
design is a $14 \times 16,384$ array, each row being a vector of 0's and 1's: a 0 in position $i$ of row $k$ means that, in the
$k$-th tree computation, gene number $i$ (in alphabetic order) were discarded, and conversely it was considered if the coefficient is 1.
Rows are thus the ``observations'' while columns correspond to regressors. 
The  1-d endogenous response variable, for its part, was a vector of size $16,384$, having an 1 in position $k$ if and only if Topology~1 
has been produced with the choice of genes corresponding to the row number $k$ in the exogenous design (resp. Topology 0, 2, or 3 in the three
other binary choice logit models). The model has been fitted using maximum likelihood with Newton-Raphson solver. Convergence has been obtained
after 8 iterations, and the Logit regression results are summarized in Table~\ref{tab:logit}.

A first conclusion of the results obtained when investigating the impact of each gene on the most supported topology is that
all considered coding sequences bring information, except perhaps the particular case of \emph{nad4} (see column $P>|z|$). 
Additionally, when the effect of a mitochondrial coding sequence is negative regarding Topology~1, its impact is not very pronounced, while \emph{cob}, \emph{cox1}, and \emph{nad5} contribute the most to this topology (see coef column: large absolute value means large effect, while negative coefficients tend to break the topology). All these findings are coherent with the frequency of occurrences of each gene in the choice 
of Topology~1: \emph{nad5}, \emph{cox1}, and \emph{nad5} were present in 12,642 computations leading to this topology (77.07\%), while only $8,685$ computations with \emph{rrnS}, \emph{nad3}, and \emph{nad1} have led to this topology (53\%), as described in Table~\ref{tab:frequences}.

Further investigations of the role of each sequence and their effects on each topologies are provided in Tables~\ref{tab:logit0}, \ref{tab:logit2}, and \ref{tab:logit3} of supplementary data, which contain the results of the dummy logit regression test for Topologies 0, 2, and 3 respectively.

\section{Conclusion}

Deep investigation of the molecular phylogeny of the \emph{Taenia} genus has been
performed in this paper. 14 coding sequences, taken from mitochondrial genomes, 
have been considered for maximum likelihood phylogenetic reconstruction. As the obtained tree was 
not satisfactorily supported, each combination from 1 to 14 genes has been further investigated, leading to $16,384$ trees representing 131 topologies. Four close topologies were then isolated whose 
differences are located in the position of \emph{T.~hydatigena} and the sister relationship between
\emph{T.~laticollis} and \emph{T.~pisiformis}. Using the logit model 
we have finally proven that Topology~1 was the most probable one and have emphasized the negative role of some genes for that phylogeny.

In future work, the authors intend to use LASSO test for regressing the bootstrap on the genes. Furthermore, we will investigate the phylogeny of \emph{Echinococcus} using
a similar approach. Indeed, there is no general agreement regarding the phylogeny of this genus. In particular, some species were discovered to have contradictory positions in the available  literature. All the possible combinations of the 12 mitochondrial genes, plus \emph{rrnL} and \emph{rrnS} and also 5 nuclear genes, will be considered, leading to the production of $43,796$ phylogenetic trees. Their topologies will be
compared, and the influence of each gene on these topologies will be rigorously measured, in order
to determine the most probable phylogenetic tree of this species. Finally, the phylogeny of the class \emph{Eucestoda}  will be investigated using a similar approach.

\bigskip

\emph{All computations have been performed using the Mésocentre de Calcul de Franche-Comté facilities.}

\bibliographystyle{plain}
\bibliography{mabase}

\section{Appendices}
\begin{table}[h]
\caption{Dummy logit regression results for Topology 0}
\begin{tabular}{|c|c|c|c|c|c|}
\hline
\textbf{} & \textbf{coef} & \textbf{std err} & \textbf{z} & \textbf{$P>|z|$} & \textbf{[95.0\% Conf. Int.]} \\ \hline
\textit{atp6 } & -1.0959 & 0.069 & -15.901 & 0.000 & [-1.231, -0.961] \\ \hline
\textit{cob } & -1.7306 & 0.073 & -23.593 & 0.000 & [-1.874, -1.587] \\ \hline
\textit{cox1 } & 0.233 & 0.066 & 3.535 & 0.000 & [0.104, 0.362] \\ \hline
\textit{cox2 } & -0.033 & 0.066 & -0.501 & 0.616 & [-0.162, 0.096] \\ \hline
\textit{cox3 } & 1.4431 & 0.071 & 20.327 & 0.000 & [1.304, 1.582] \\ \hline
\textit{nad1 } & -2.6491 & 0.082 & -32.159 & 0.000 & [-2.811, -2.488] \\ \hline
\textit{nad2 } & -5.767 & 0.151 & -38.171 & 0.000 & [-6.063, -5.471] \\ \hline
\textit{nad3 } & -0.0797 & 0.066 & -1.211 & 0.226 & [-0.209, 0.049] \\ \hline
\textit{nad4 } & 2.4925 & 0.08 & 31.017 & 0.000 & [2.335, 2.650] \\ \hline
\textit{nad4l } & -0.0296 & 0.066 & -0.449 & 0.653 & [-0.159, 0.099] \\ \hline
\textit{nad5 } & -2.5196 & 0.081 & -31.123 & 0.000 & [-2.678, -2.361] \\ \hline
\textit{nad6 } & -1.0355 & 0.069 & -15.097 & 0.000 & [-1.170, -0.901] \\ \hline
\textit{rrnL } & -1.403 & 0.071 & -19.803 & 0.000 & [-1.542, -1.264] \\ \hline
\textit{rrnS } & 2.2175 & 0.078 & 28.594 & 0.000 & [2.066, 2.370] \\ \hline
\end{tabular}
\label{tab:logit0}
\end{table}

\begin{table}[h]
\caption{Dummy logit regression results for Topology 2}
\begin{tabular}{|c|c|c|c|c|c|}
\hline
\textbf{} & \textbf{coef} & \textbf{std err} & \textbf{z} & \textbf{$P>|z|$} & \textbf{[95.0\% Conf. Int.]} \\ \hline
\textit{atp6 } & 0.3534 & 0.055 & 6.479 & 0.000 & [0.247, 0.460] \\ \hline
\textit{cob } & -0.3845 & 0.055 & -7.042 & 0.000 & [-0.492, -0.277] \\ \hline
\textit{cox1 } & -1.5321 & 0.059 & -25.903 & 0.000 & [-1.648, -1.416] \\ \hline
\textit{cox2 } & -1.0754 & 0.057 & -18.956 & 0.000 & [-1.187, -0.964] \\ \hline
\textit{cox3 } & -1.7371 & 0.06 & -28.729 & 0.000 & [-1.856, -1.619] \\ \hline
\textit{nad1 } & 2.305 & 0.065 & 35.601 & 0.000 & [2.178, 2.432] \\ \hline
\textit{nad2 } & 3.6525 & 0.078 & 46.902 & 0.000 & [3.500, 3.805] \\ \hline
\textit{nad3 } & 0.8116 & 0.056 & 14.577 & 0.000 & [0.702, 0.921] \\ \hline
\textit{nad4 } & -1.4916 & 0.059 & -25.323 & 0.000 & [-1.607, -1.376] \\ \hline
\textit{nad4l } & -1.641 & 0.06 & -27.427 & 0.000 & [-1.758, -1.524] \\ \hline
\textit{nad5 } & -3.4317 & 0.075 & -45.481 & 0.000 & [-3.580, -3.284] \\ \hline
\textit{nad6 } & 0.2363 & 0.054 & 4.343 & 0.000 & [0.130, 0.343] \\ \hline
\textit{rrnL } & -0.7259 & 0.055 & -13.1 & 0.000 & [-0.834, -0.617] \\ \hline
\textit{rrnS } & -1.3261 & 0.058 & -22.878 & 0.000 & [-1.440, -1.213] \\ \hline
\end{tabular}
\label{tab:logit2}
\end{table}

\begin{table}[h]
\caption{Dummy logit regression results for Topology 3}
\begin{tabular}{|c|c|c|c|c|c|}
\hline
\textbf{} & \textbf{coef} & \textbf{std err} & \textbf{z} & \textbf{$P>|z|$} & \textbf{[95.0\% Conf. Int.]} \\ \hline
\textit{atp6 } & -0.9133 & 0.081 & -11.331 & 0.000 & [-1.071, -0.755] \\ \hline
\textit{cob } & -1.3941 & 0.085 & -16.49 & 0.000 & [-1.560, -1.228] \\ \hline
\textit{cox1 } & -0.1349 & 0.078 & -1.739 & 0.082 & [-0.287, 0.017] \\ \hline
\textit{cox2 } & 0.8051 & 0.08 & 10.12 & 0.000 & [0.649, 0.961] \\ \hline
\textit{cox3 } & -0.7384 & 0.08 & -9.278 & 0.000 & [-0.894, -0.582] \\ \hline
\textit{nad1 } & -2.0258 & 0.092 & -22.074 & 0.000 & [-2.206, -1.846] \\ \hline
\textit{nad2 } & -0.2326 & 0.078 & -2.992 & 0.003 & [-0.385, -0.080] \\ \hline
\textit{nad3 } & -0.7489 & 0.08 & -9.406 & 0.000 & [-0.905, -0.593] \\ \hline
\textit{nad4 } & -3.5617 & 0.131 & -27.208 & 0.000 & [-3.818, -3.305] \\ \hline
\textit{nad4l } & 0.4031 & 0.078 & 5.168 & 0.000 & [0.250, 0.556] \\ \hline
\textit{nad5 } & 2.1308 & 0.092 & 23.219 & 0.000 & [1.951, 2.311] \\ \hline
\textit{nad6 } & -1.1314 & 0.082 & -13.766 & 0.000 & [-1.292, -0.970] \\ \hline
\textit{rrnL } & -3.8926 & 0.146 & -26.68 & 0.000 & [-4.179, -3.607] \\ \hline
\textit{rrnS } & 0.2623 & 0.078 & 3.376 & 0.001 & [0.110, 0.415] \\ \hline
\end{tabular}
\label{tab:logit3}
\end{table}

\end{document}